\documentclass[runningheads]{llncs}
\usepackage{graphicx}
\usepackage{multirow}
\usepackage{comment}
\usepackage{float}
\usepackage{subfig}
\usepackage{color}
\usepackage{amsmath}

\begin{document}
\title{MSU-Net: Multiscale Statistical U-Net for Real-time 3D Cardiac MRI Video Segmentation} 
\titlerunning{MSU-Net}
%
\author{Tianchen Wang\inst{1}
\and Jinjun Xiong\inst{2}
\and Xiaowei Xu\inst{1}
\and Meng Jiang\inst{1}
\and Haiyun Yuan\inst{3} 
\and Meiping Huang\inst{3} 
\and Jian Zhuang\inst{3}
\and Yiyu Shi\inst{1}
}
%
\authorrunning{T. Wang, et al.}
\institute{University of Notre Dame, \\
\email{\{twang9, xxu8, yshi4\}@nd.edu}
\and IBM Thomas J. Watson Research Center, \\ 
\email{jinjun@us.ibm.com}
\and Guangdong General Hospital\\
\email{yhy\_yun@163.com, huangmeipng@126.com, zhuangjian5413@tom.com}}

\maketitle              
\begin{abstract}
Cardiac magnetic resonance imaging (MRI) is an essential tool for MRI-guided surgery and real-time intervention. The MRI videos are expected to be segmented on-the-fly in real practice. However, existing segmentation methods would suffer from drastic accuracy loss when modified for speedup. In this work, we propose Multiscale Statistical U-Net (MSU-Net) for real-time 3D MRI video segmentation in cardiac surgical guidance. Our idea is to model the input samples as multiscale canonical form distributions for speedup, while the spatio-temporal correlation is still fully utilized. A parallel statistical U-Net is then designed to efficiently process these distributions. The fast data sampling and efficient parallel structure of MSU-Net endorse the fast and accurate inference. Compared with vanilla U-Net and a modified state-of-the-art method GridNet, our method achieves up to 268\% and 237\% speedup with 1.6\% and 3.6\% increased Dice scores.
\end{abstract}

\section{Introduction}
Real-time Magnetic Resonance Imaging (MRI) techniques have been providing fast and accurate visual guidance in multiple fields.
The duration of cardiac surgery (e.g., prosthetic valve implantation in the correct location at the aortic annulus) has been significantly shortened since interactive real-time MRI was getting applied \cite{mcveigh2006real}.
Interventional real-time MRI has also been adopted for congenital, ischemic, and structural heart disease for its capacity of visualizing 3D anatomy and assessing myocardial tissue as well as local hemodynamics \cite{campbell2017real}.
To achieve real-time MRI guidance, the images need to be segmented on-the-fly, at a speed of at least \textbf{30}, preferably up to 100 frames per second (FPS) \cite{schaetz2017accelerated,iltis2015high}.

However, performing real-time segmentation on cardiac MRI images is a challenging task.
In addition to the difficult effects such as anisotropic resolution, cardiac border ambiguity and large variations among targeting objects from patients \cite{zheng20183d}, the requirement of real-time fast segmentation demands a lightweight and efficient processing framework.
Existing approaches used complicated neural network architectures to achieve good accuracy and were not able to make inference in real time \cite{isensee2017automatic,zotti2018convolutional}.
Recently, Statistical Convolutional Neural Network (SCNN) was proposed to speed up conventional CNNs with little performance loss in video object detection \cite{wang2019scnn}.
Instead of feeding the input samples as deterministic values, SCNN used Independent Component Analysis (ICA) 
to extract parameterized statistical distributions in canonical form to compactly model the temporally and contextually correlated information. Then the network model propagated the distributions in canonical form more efficiently than deterministic values.

In this work, we propose Multiscale Statistical U-Net (MSU-Net) for real-time cardiac MRI segmentation. We incorporate SCNN and a new multiscale data sampling method with the U-Net to capture spatio-temporal correlation in the input data. 
Our model adopts a parallel architecture to efficiently propagate the multiscale distributions.
Specifically, we apply ICA with multiple sets of temporal image patches to generate a cluster of canonical form distributions, each of which represents a different scale to model the input data. 
This multiscale sampling method can preserve the information of spatio-temporal correlations at different scales. 
Then we implement a number of parallel yet light-weight encoder-decoder style branches for efficient inference. 
Each branch propagates the specific scale of canonical form distributions. 
Experimental results show that our MSU-Net achieves up to 268\% and 237\% speedup with 1.6\% and 3.6\% increased Dice scores compared with vanilla U-Net and a modified state-of-the-art method GridNet \cite{zotti2018convolutional}.

\section{Background}
SCNN \cite{wang2019scnn} was the first model that feeds CNNs with a reasonable number of statistical distributions that were decomposed from the input data. SCNN is lighter and thus of higher speed than conventional CNNs that conduct deterministic operations (such as sum and max).

SCNN applied ICA to decompose video frames that exhibit spatio-temporal correlation into canonical form distributions as follows:
\begin{equation}
D\!=\!a_{0}\!+\!a_{1}X_1\!+\!\dots\!+\!a_{m}X_m\!+\!a_{r}R,
\label{eq:icar}
\end{equation}
where (1) $D$ is a random multivariate signal, which in video object detection represents the same pixel across multiple frames in a snippet; (2) $a_{0}$ is the mean value of $D$; (3) $X_{i}~(i\in\{1 \dots m\})$ are additive independent subcomponents of $D$; (4) $a_{i}~(i\in\{1 \dots m\})$ are the corresponding weight act as mixing matrix; (5) $R$ denotes uncorrelated Gaussian noise and $a_{r}$ is the weight of $R$; (6) $m$ is the basis dimension of the canonical form distribution. 

With the help of predefined core operations (weighted {\em sum} and {\em max}) that keep their outputs still in canonical form distributions, 
SCNN needs little modification to the standard gradient descent based scheme. 
It can be trained using the same forward and back propagation procedures as conventional CNNs. At the output, the results are mixed to form a temporal feature map for each sample by plugging in the values of independent sources $X_i$ from the ICA process. 
By processing multiple frames at a time through distributions, SCNN significantly speedups object detection in videos over conventional CNNs with slight accuracy degradation.

\section{Method}
In this section, we first present a multiscale sampling method to extract canonical form distributions from input 3D MRI videos.
Then we introduce the architecture of MSU-Net and explain how it processes these distributions for real-time segmentation.

\subsection{Distribution Extraction with Multiscale Data Sampling}
\label{sec:corr_model}
In order to build linear distributions in parameterized canonical form (Equation~\ref{eq:icar}) via ICA, 
we need to decide how to properly extract samples from 3D MRI video to feed into ICA, i.e., what information each $D$ should represent. 
In the approach of SCNN for video object detection \cite{wang2019scnn}, the video clips are resized and split into small snippets, and each distribution $D$ models the same pixel across multiple frames in the same snippet. 
However, this cannot be applied to 3D MRI video directly since lots of semantic details important to segmentation would be lost. 
Thus, we propose to use $D$ to represent a patch within a small range (both spatially and temporally) where strong correlation exists. Specifically, 
we denote the dimension of an input 3D MRI video as [$X$, $Y$, $Z$, $T$], where $X-Y$ plane is the short axis plane, $Z$ is the short axis and $T$ is the temporal dimension.
The common issues of slice shifting as well as large inter-slice gap in MRI cardiac images along short-axis ($Z$ axis) \cite{zotti2018convolutional} lead to minimum spatial correlation in $X-Z$ and $Y-Z$ planes.
Therefore, we extract patches within the dimension [$X$, $Y$, $T$], independent of $Z$.

\begin{figure}[!t]
\begin{center}
\includegraphics[width=1\linewidth]{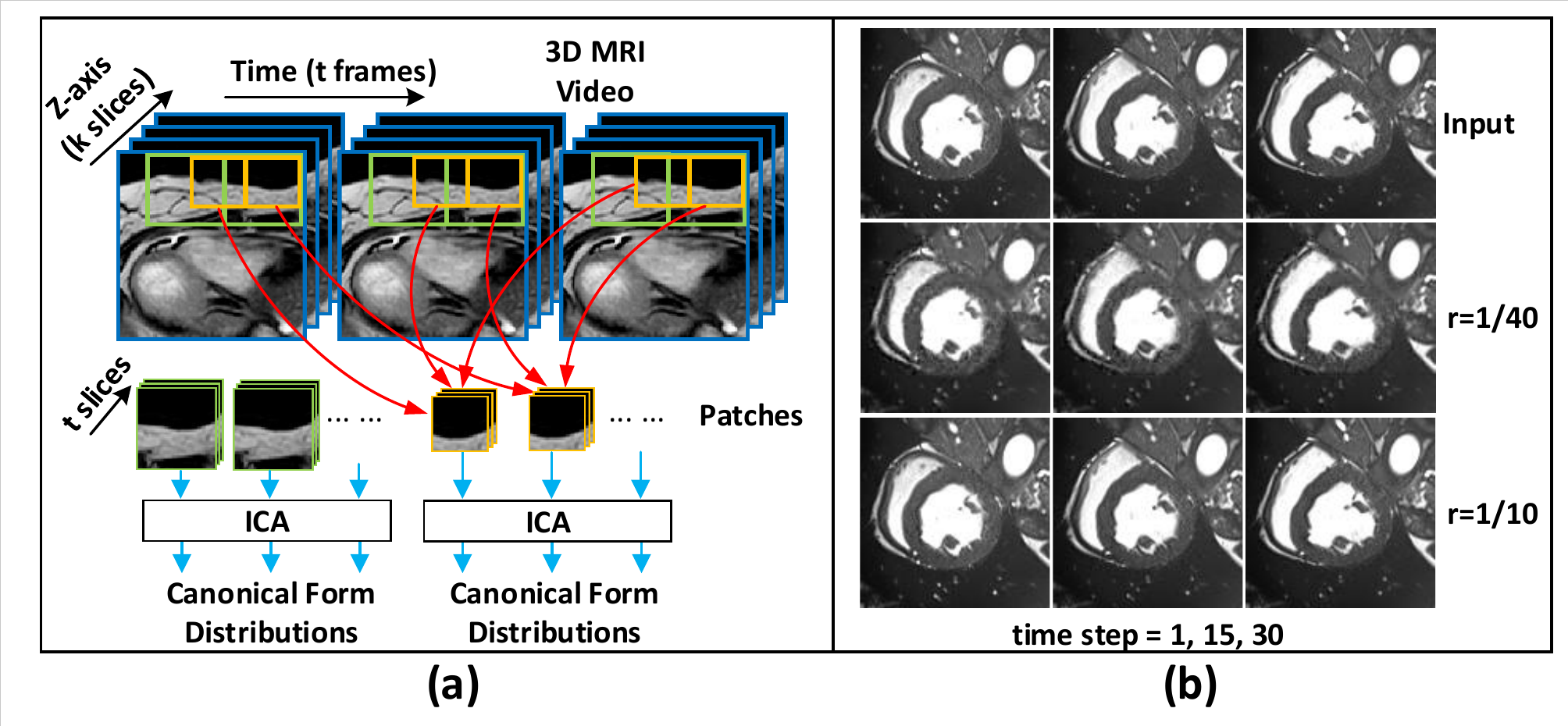}
\end{center}
\caption{(a) Illustration of multi-scale data sampling from cropped 3D MRI video. Each canonical form is extracted using the samples from the same position in the $X-Y$ plane and collected at different time steps. Different canonical forms can have different patch sizes. (b) Restored inputs with multi-scale data sampling method
at different time steps (t) using different compression ratio (r). 
The restored inputs with r=1/40 have more noises than those with r=1/10.
}
\label{fig:cropping_restore}
\end{figure}

Before extracting the patches, the 3D MRI videos are normalized first to remove offsets among videos. Each patch is then extracted using a window of size $(n, n)$ on the $X-Y$ plane over $t$ time steps. 
We call $t$ as snippet span. We propose to allow different canonical forms to have different $n$ and $t$, as such an approach covers potential spatio-temporal correlations at different scales. We call this cluster of distributions with multiple patch sizes as multiscale distributions.
An example of the extraction process of multiscale distributions with different patch sizes on one slice is shown in Fig.~\ref{fig:cropping_restore} (a).
The patches are collected at the same position over time and fed to ICA to extract canonical form distributions.
ICA has to be used because the propagation of the canonical form distributions requires all the bases to be independent. 
Other approahces such as PCA cannot guarantee this unless the samples follow Gaussian distributions, which is not the case in our problem. 
As a result, the snippet of 3D MRI video is ``collapsed'' into a smaller 3D image, with each voxel representing a canonical form distribution (Equation~\ref{eq:icar}) that has both spatial and temporal correlations: with patch size $(n, n, t)$ and predetermined independent basis dimension $d$, a compression ratio of  $r=d/(n^2t)$ is achieved.

To show the feasibility of the proposed data sampling, we extract the multiscale canonical form distributions using our procedure with various compression ratio $r$ by changing the basis dimension $d$ with $n=7$ and $t=5$. 
The visual results along with the compression ratio are shown in Fig.~\ref{fig:cropping_restore} (b).
With a larger ratio ($\text{r}\!=\!1/40$, smaller basis dimension of canonical form distribution), the restored video gain more noise with vague contours, which would bring obstruction to the segmentation task.
With a smaller ratio ($\text{r}\!=\!1/10$), the difference between the input video and the restored mixing video is negligible.
Therefore, we adopt $\text{r}\!=\!1/10$ as the compression ratio in our following experiments.

\begin{figure}[!t]
\begin{center}
\includegraphics[width=1\linewidth]{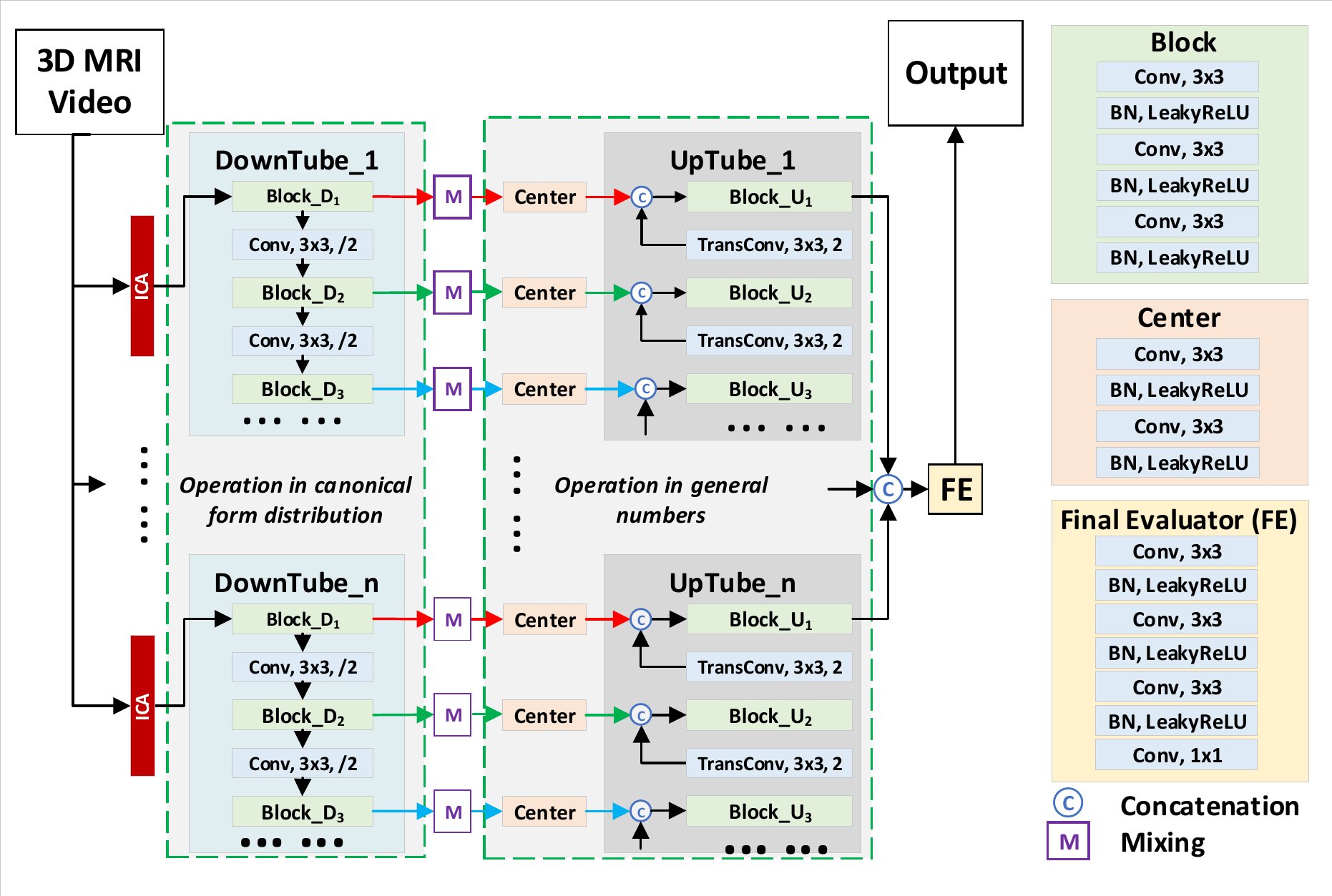}
\end{center}
\caption{The architecture of MSU-Net. The number of Blocks in DownTube/UpTube varies to accommodate the various input dimensions.}
\label{fig:msunet}
\end{figure}

\subsection{Real-time Segmentation with MSU-Net}
\label{sec:msunet}
The multiscale canonical form distributions provide compact data representation for efficient processing. In this subsection, we explore a parallel structure, namely MSU-Net, that can further speedup the segmentation.

Fig.~\ref{fig:msunet} illustrates our MSU-Net which consists of multiple DownTubes (DTs), UpTubes (UTs), Center blocks, and a final evaluator (FE).
The DTs and UTs act as the encoders and decoders in U-Net for feature propagation.
Multiple DTs are built for a set of splitting patch sizes, each consisting of multiple blocks with downscaling convolution layers to perform feature downscaling and reuse. 
The ICA process and the corresponding mixing operations are done before and after the operations in DT, repsectively, and the operations in DT are performed in canonical form distributions similar to the work developed in SCNN~\cite{wang2019scnn}. 
The features in UT are propagated and upscaled with the blocks made of convolutional layers, and transposed convolutional layers, respectively.
The features after each upscaling are concatenated with the one skipped from DT for feature reuse. 
After the outputs are obtained from UTs with various patch sizes, all features would have the same dimensions, which are then concatenated and forwarded to the final evaluator to generate the final output. 
The number of blocks in DT/UT varies to accommodate the input dimensions of 3D images.

\section{Experiments} 
\subsection{Experiment Setup}
The evaluation task is to segment right ventricle (RV), myocardium (MYO), and left ventricle (LV) from MRI video clips in real time. 
We evaluate the proposed MSU-Net and competitive baselines on segmenting the RV, MYO and LV from the frames of End Diastolic (ED) and End Systolic (ES) instant. 
These frames were collected from the ACDC MICCAI 2017 challenge dataset \cite{bernard2018deep} with additional labeling done by experience radiologists. These frames have similar properties as 3D cardiac MRI videos.
The dataset has 150 exams from different patients with 100 for training and 50 for testing.
The images were collected following the common clinical SSFP cine-MRI sequence with a series of short-axis slices starting from the mitral valves down to the apex of the left ventricle.
We perform 5-fold cross-validation and use the Dice score to evaluate the segmentation accuracy. 

We implement two versions of MSU-Net with specific snippet spans ($\text{t}=5$ and $10$, denoted as T5, T10, respectively) for evaluation.
The ICA processing time is included when we evaluate the inference time of MSU-Net. 
Existing approaches have reported their FPS on the same dataset: 
$\sim1$ \cite{isensee2017automatic}, 
and 
$5.56$\cite{zotti2018convolutional}. 
Clearly, none of them can perform real-time inference (i.e., at least 30 FPS). 
Therefore, we modify and rebuild these approaches to speed them up so that they can be compared with MSU-Net on a relatively fair basis. 
We implement a set of shallower/slimmer versions (i.e., with fewer layers/fewer channels) of the models. 
Specifically, we modify the 2D U-Net \cite{ronneberger2015u} to a shallow version with a depth of 3 and initial filter of 8 or 16. 
We denote them as D3+IF8 and D3+IF16, respectively. 
The 2D U-Net with vanilla configuration (D5+IF64) is also included. 
We also modify GridNet to shallower versions with a depth of 2 or 3 and initial filter of 32, denoted as D2+IF32 and D3+IF32, respectively. 
The vanilla version of GridNet is one of the best models in the ACDC 2017 challenge. 
All these methods are fully trained after modification. 

In our experiments, we do not include 3D U-Net for comparison due to its excessive memory consumption, unbalanced input dimensions, and slow inference speed \cite{ronneberger2015u,isensee2017automatic}.
Meanwhile we do not include lightweight networks such as ShuffleNet \cite{ma2018shufflenet}
or MobileNet \cite{sandler2018mobilenetv2} 
which is designed with small memory footprint for mobile devices in image classification/object detection rather than medical segmentation, while inference speed is not their primary concern (which mainly depends on the network depth).
We have tried ShuffleNet/MobileNet in our experiment settings and the speeds are only at 8.45/11.58 FPS, which are slower than the nets we reported.

We implement MSU-Net and 2D U-Nets using PyTorch.
The GridNet was implemented using TensorFlow \cite{zotti2018convolutional}. 
All experiments run on a machine with 16 cores of Intel Xeon E5-2620 v4 CPU, 256G memory, and an NVIDIA GeForce GTX 1080 GPU.

\begin{table}[!tb]
\centering
\caption{Comparison between baseline methods and our MSU-Net on Dice score and FPS for 3D MRI video segmentation. ``T5''/``T10'' denotes the video snippet span in MSU-Net (t=5/t=10). ``D'' and ``IF'' denote the depth of the network and the initial filters number of the input layer, respectively.
}
\begin{tabular}{|l||c||c|c|c|c|}
\hline
\multirow{2}{*}{\textbf{Methods}} & \multirow{2}{*}{\textbf{FPS}} &  \multicolumn{4}{|c|}{\textbf{Dice score}}\\ \cline{3-6}   
& & \textbf{RV} & \textbf{MYO} & \textbf{LV}  & \textbf{Average}\\ \hline \hline
GridNet (D3+IF32) & 15.7 & .842$\pm$.028 & .804$\pm$.026 & .901$\pm$.036 & .849$\pm$.014 \\
U-Net (D5+IF64) & 16.1 & \textbf{.865$\pm$.036} & .761$\pm$.039 & \textbf{.911$\pm$.026} & .846$\pm$.025\\
GridNet (D2+IF32) & 18.2 & .815$\pm$.025  & .812$\pm$.014 & .851$\pm$.033 & .826$\pm$.011 \\
U-Net (D3+IF16) & 33.2 & .564$\pm$.071 & .738$\pm$.045 & .767$\pm$.026 & .690$\pm$.036 \\
U-Net (D3+IF8) & 43.2 & .552$\pm$0.079 & .674$\pm$.060 & .759$\pm$.059 & .662$\pm$.058 \\ \hline
MSU-Net (T5) & \textbf{43.2} & .855$\pm$.026 & \textbf{.836$\pm$.022} & .897$\pm$.017 & \textbf{.862$\pm$.011} \\
MSU-Net (T10) & \textbf{70.2} & .837$\pm$.034 & .811$\pm$.049 & .854$\pm$.040 & .834$\pm$.020 \\ 
\hline
\end{tabular}
\label{table:compare}
\end{table}

\subsection{Results}
Table~\ref{table:compare} presents the comparison among U-Net, GridNet, and the proposed MSU-Net on Dice score and FPS.
Our MSU-Nets can achieve the fastest processing speed (highest FPS) and the best Dice score.
Compared with the fastest baseline method U-Net (D3+IF8), our MSU-Net (T10) runs 1.63$\times$ faster and makes an improvement of 26\% on segmentation accuracy. 
Compared with the most accurate baseline method (with the highest Dice score) GridNet (D3+IF32), our MSU-Net (T5) can achieve a slightly higher accuracy and 2.75$\times$ faster processing speed. From the table, it is clear that MSU-Nets are the only capable method to segment real-time 3D MRI videos. 

For MSU-Net, a bigger video snippet span (T10) can obtain a faster processing speed with a slight accuracy degradation (only 0.028).
However, for U-Net, when it is modified into shallow/slim versions such as U-Net (D3+IF16) and (D3+IF8) for real-time processing ($\ge 30$ FPS), the accuracy degrades significantly: We observe that the accuracy drops from 0.846 to 0.690 and 0.662, respectively.
We observe the same pattern for GridNet, and conclude that MSU-Net can achieve a stable accuracy when configured for segmentation in real time.

\begin{figure}[!t]
\begin{center}
\includegraphics[width=1\linewidth]{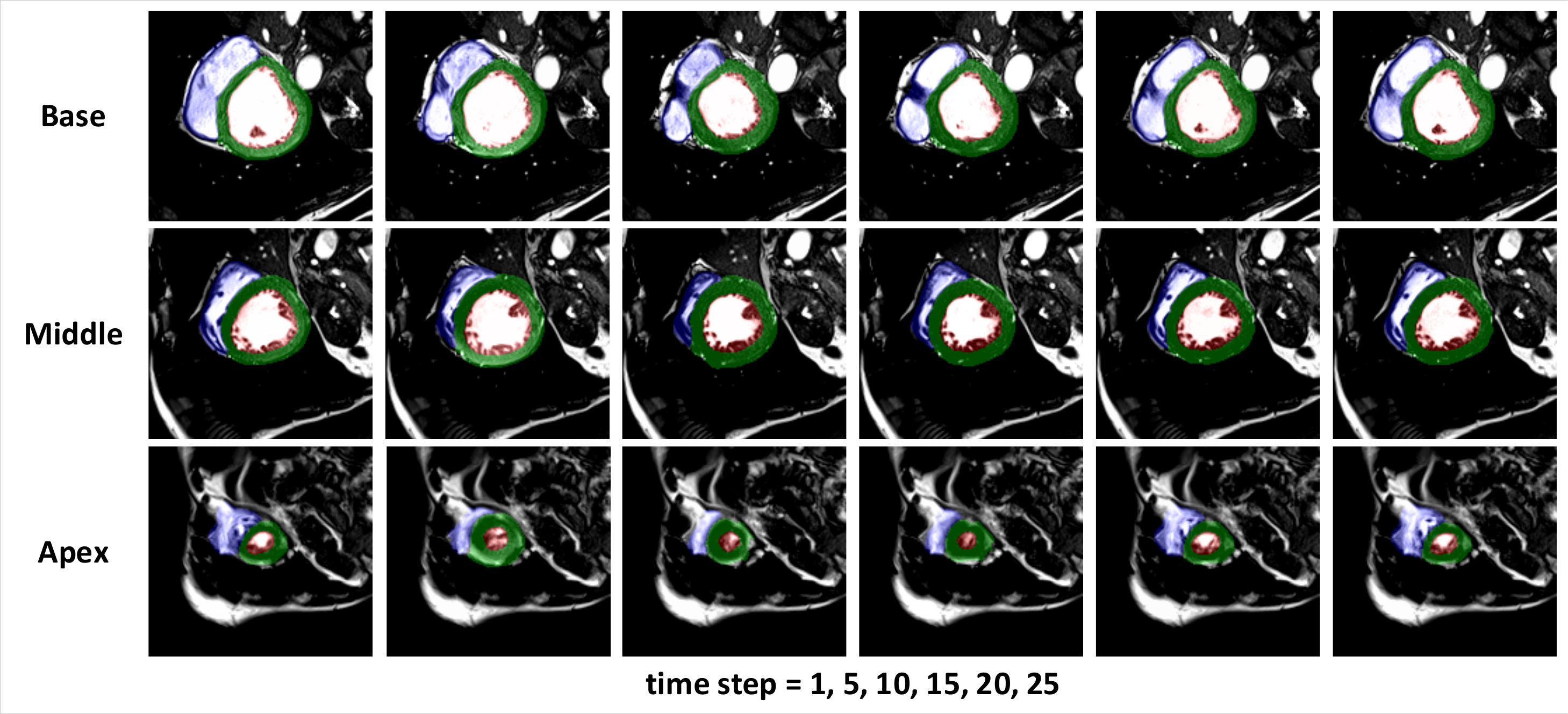}
\end{center}
\caption{The segmentation results of our method MSU-Net (T5) on the testing data. The rows indicate the slices at the base, the middle, and the apex of LV. The columns show the results at various time steps in series. 
RV, MYO, and LV are labeled in blue, green and red, respectively. }
\label{fig:seg_out}
\end{figure}

Finally, Fig.~\ref{fig:seg_out} shows the examples of MSU-Net segmentation results at various time steps.
Note that our MSU-Net can accurately segment the target areas. The boundaries are clearly extracted
on most of the slices.
In the base and middle slices, the segmentation fits the contours of targets.
In some of the apex slices, the segmentation of RV (labeled in blue) is not as accurate as MYO and LV, because of the unclear boundaries between the instances.

\section{Conclusions}
In this paper, we proposed Multiscale Statistical U-Net (MSU-Net) for real-time 3D cardiac MRI video segmentation. Based on the scheme of Statistical Convolutional Neural Network, we model the input samples 
as multiscale canonical form distributions for speedup, while the spatio-temporal correlationis still fully utilized. A parallel statistical U-Net is then proposed to process these multiscale distributions 
efficiently. On the 3D cardiac MRI videos from the ACDC MICCAI 2017 dataset, MSU-Net achieves up to 268\% and 237\% speedup with 1.6\% and 3.6\% increased Dice scores compared with vanilla U-Net and a modified state-of-the-art method GridNet, respectively.

\section{Acknowledgement}
This work was approved by the Research Ethics Committee of Guangdong General Hospital, Gunagdong Academy of Medical Sciences with protocol No. 20140316. This work was supported by the National key Research and Development Program [2018YFC1002600], Science and Technology Planning Project of Guangdong Province, China [No. 2017A070701013, 2017B090904034, 2017030314109, and 2019B020230003], National Science Foundation grant [CCF-1919167], and Guangdong peak project [DFJH201802].

\bibliographystyle{splncs04}
\bibliography{bib}

\end{document}